\documentstyle[]{article}
\font\cero=cmss10 scaled 1728 \font\uno=cmssbx10 scaled 1200
\begin{document}
\small{
\begin{flushleft}
{\cero  Poincar\'e charges for chiral membranes} \\[3em]
\end{flushleft}
{\sf Alberto Escalante}\\
{\it  Departamento de f\'{\i}sica, Centro de investigaci\'on y de
Estudios Avanzados del I.P.N.,\\}
Apartado Postal 14-740, 0700 M\'exico, D.F., M\'exico \\ (aescalante@fis.cinvestav.mx) \\[4em]
\noindent{\uno Abstract} \vspace{.5cm}\\
Using basic ideas of simplectic geometry, we find the covariant canonically conjugate variables, the commutation relations and the Poincar\'e charges for chiral superconducting membranes (with null  currents ), as well as we find the stress tensor for the theory under study.\\
\noindent \\

\begin{center}
{\uno I. INTRODUCTION}
\end{center}
\vspace{1em} \ The interest in physical systems characterized by
extended structures goes back to the XIX th Century and to Lord
Kelvin's "aether atoms'', for which a spatial extension was
postulated in order to accommodate a complex structure which would
be have both as
 an elastic solid (conveying the transverse wave motion of electromagnetism)
 and viscous liquid (dragged by the earth in its orbital motion).\\
In the XX  Century, there have been three active motivations
leading to either classical or quantum extendons. On the other
hand, the physics of condensed matter  (including biological
systems) have revealed that membranes and two-dimensional layers
play an important role; in some case, there also appear
one-dimensional filaments (or strings). Similar structures appear
in astrophysics and cosmology, one example being the physics of
Black holes, in which the "membrane" is the  boundary layer
between the hole and the embedding spacetime, and another example
is represented by the hypothetical cosmic strings \cite{1a}.\\
In this context, it is believed  that cosmic strings are fundamental bridges on the understanding of the  Universe formation due to  that several cosmological phenomena can be described by means of cosmic strings properties.  Besides of these there are other kinds of cosmic objects possessing different properties  of those inherited to ordinary cosmic strings, for  example: domains  walls and  hybrid structures.  They can  arise in several Grand Unified  Theories whenever  there exists an appropriate symmetry  breaking scheme. However,  there is other class of cosmological objects  that can emerge
with  ability  to carry  some sort of charge.  For instance, as was suggested by Witten \cite{1},  cosmic strings could behave like superconductors. \\
Since  that  time,  the vast research  on super conducting strings has thrown a new variety   of cosmic objects. The cosmological results of supersymmetric theories  was also considered, yielding  to another class of cosmic strings, namely chiral cosmic strings. These objects   are the result of symmetry breaking in supersymmerty (SUSY)  where a $U(1)$ symmetry is broken with  a Fayet-lliopoulos $D$ term, turning out a sole fermion zero mode traveling  in only one direction  in the string core \cite{2}. In other words, when the current along  the superconducting string  shows a light-like causal structure then we have a chiral string. The  dynamics  of the chiral strings model has been recognized to be an intermediate stage 
between  Dirac-Nambu-Goto  [DNG] model and that of generic elastic model \cite{3,4,5}, which  has interesting cosmological  implications. The microphysics of this kind  of topological defects has been investigate, opening up the possibility to have chiral  vortons  more stable  than vortons  of other kinds.\\
In this manner, the purpose of this article is establish the basis for the study of symmetries   and the quantization aspects of chiral membranes.  Such basis consists in the construction of a covariant and gauge invariant symplectic structure on the corresponding quotient phase space $Z$ (the space of solutions of classical equations of motion divided by the symmetry group volume) instead of choosing a spatial coordinate system on the phase space. For example, recent letters \cite{6,7,8,9,10} the basic elements to quantize extended objects (in particular bosonic p-branes) has been
explored;  in \cite{7} we established  the basis to
study the quantization aspects of p-branes with thickness,
because, when we add it to the [DNG] action has an important
effect in QCD \cite{11}, among other things. In \cite{8} has been
demonstrated that the presence of Gauss-Bonnet [GB] topological
term in the [DNG] action describing strings, has a dramatic effect
on the covariant phase space formulation of the theory, in this
manner, we shall obtain a completely different quantum field
theory. Recently, using the results given in \cite{8} we identify
the covariant canonical variables for [DNG] p-branes and [GB]
strings, among other things \cite{9}.  Thus, in this paper we  extend the   results  for chiral strings reported in \cite{12,13, 14a},   using  a Kaluza-Klein (KK) reduction mechanism  \cite{14} and following  closely  the results given in \cite{7,8,9}. \\
This paper is organized as follows. In Sect. II, We discuss  the deformation  formalism  for the geometry of chiral membranes, which is crucial for the develop of this work. In Sect. III, using the results given in \cite{8,9} and basic ideas of symplectic geometry \cite{15}, we identify the covariant canonically conjugate variables, the relevant Poison brackets,   the Poncar\'e charges and its corresponding laws of conservation. In the Sect. IV, we study a generalization of auxiliary  variables  for relativistic membranes for an arbitrary co-dimension , which is a generalization of the results given in \cite{16}. We find in this section the stress tensor for the theory under study and confirm the results given in the last section. In Sect. V, we establish some remarks and prospects. 
\setcounter{equation}{0} \label{c2}
\newline
\newline
\newline
\noindent \textbf{II. Geometry and deformations for chiral membranes}\\[1ex]
In this section, utilizing the formalism given in \cite{17} we will
discuss the deformations of the embedding for chiral membranes possesing nullcurrents on the worldvolume  $(\varpi= \gamma^{ab} \varphi,_a \varphi,_b)$ based in the Kaluza-Klein approach used in \cite{14} . For ours  purposes , we consider a relativistic membrane of dimension $d$, whose worldsheet $\{m, \Gamma _{ab} \}$  is  an oriented timelike $d+1$ dimensional manifold embedded in a $N$-dimensional extended arbitrary fixed background space-time $\{ M, g_{\bar{\mu} \bar{\nu }} \}$. We describe the worldsheet by the extended embedding as
\begin{equation}
X^{\bar{\mu}}= \left(
\begin{array}{rr}
X^{\mu}(\xi^{a})   \\
\varphi(\xi^{a})  \\
\end{array}
\right),
\end{equation}
where $\varphi$ is a field living on the worldsheet,
$\bar{\mu}=0,1,...N$, and $\xi^a$ are coordinates on the
worldsheet, $a=1, 2...d$.\\
With the former embedding, we can  make contact with the
Kaluza-Klein description for the background spacetime metric
\begin{equation}
g_{\bar{\mu}\bar{\nu}}= \left(
\begin{array}{rr}
g_{\mu \nu} & 0   \\
0 & g_{44} \\
\end{array}
\right),
\end{equation}
where $g_{\mu \nu}$ is the original background spacetime and
$g_{44}$ is a constant.\\
We have for the embedding a tangent basis defined by
$e_{a}=X^{\bar{\mu}},_{a}D_{\bar{\mu}}$, where
$D_{\bar{\mu}}$ is the covariant derivative compatible with
$g_{\bar{\mu}\bar{\nu}}$. The tangent vectors
$e^{{\bar{ \mu}}}_{a}$ associated with the embedding (1) can
be written as
\begin{equation}
e^{{\bar{\mu}}}_{a}= \left(
\begin{array}{rr}
e^{\mu}_{a}   \\
\varphi,_{a}  \\
\end{array}
\right).
\end{equation}
Thus, the metric induced on $m$ is iven by $\Gamma_{ab}= g_{\bar{\mu}\bar{\nu}}e^{{\bar{\mu}}}_{a}e^{{\bar{\nu}}}_{b}= \gamma_{ab} + g_{44}\varphi,_a \varphi,_b $, where $\gamma_{ab}$ is the metric of the embedding  without the field $\phi$ \cite{17}.\\
We will denote the $n^{J \bar{\mu}}$ as the $J$-th unit
normal to the worldsheet, $I=1,...,N-d$ given by
\begin{equation}
n^{\bar{\mu}i}=\left(
\begin{array}{rr}
n^{\mu i}   \\
0  \\
\end{array}
\right), \quad \quad n^{\bar{\mu}4}= \sqrt{g_{44}} \left(
\begin{array}{rr}
e^{\mu}_{a}\varphi^{'a}   \\
- g^{44}  \\
\end{array}
\right),
\end{equation}
and defined intrinsically by 
\begin{equation}
g_{\bar{\mu} \bar{\nu}}n^{\bar{\mu} I}e^{\bar{\nu}}_{a}=0, \quad
\quad
g_{\bar{\mu}\bar{\nu}}n^{\bar{\mu}I}n^{\bar{\nu}J}=\delta{^{I}}_{J}.
\end{equation}
The collection of vectors $\{e^{\bar{\mu}}_{a}, n^{ \bar{\nu}I}\}$
can be used as a basis for the spacetime vectors appropriate for
the geometry under consideration.\\
We can calculate the world sheet gradients of the basis vectors,
given by
\begin{eqnarray}
\nonumber D_{a} e{^{\bar{\mu}}}_{b} \!\! & = & \!\!
\Gamma{_{ab}}^{c}e{^{\bar{\mu}}}_{c}-
K{_{ab}}^{I}n{^{\bar{\mu}}}_{I}, \\
D_{a}n^{\bar{\mu}I} \!\! & = & \!\! K{_{ab}}^{I} e^{b \bar{\mu}} +
\omega{_{a}}^{IJ} n{^{\bar{\mu}}}_{J},
\end{eqnarray}
where the $I$-th extrinsic curvature $K{_{ab}}^{I}$ is given by
\begin{equation}
K{_{ab}}^{I}=K{_{ba}}^{I}=-
D_{a}e{^{\bar{\mu}}}_{b}n{_{\bar{\mu}}}^{I},
\end{equation}
and the extrinsic twist potential by \cite{17}
\begin{equation}
\omega{_{a}}^{IJ}= - \omega{_{a}}^{JI}= D_{a}
n{^{\bar{\mu}I}}n{_{\bar{\mu}}}^{J}.
\end{equation}
Now we will calculate the deformations of the intrinsic and
extrinsic geometry. For we aims we consider the neighboring
surface described by the deformation
\begin{equation}
x^{\bar{\mu}}= X^{\bar{\mu}}(\xi^{a})+ \delta
X^{\bar{\mu}}(\xi^{a}),
\end{equation}
 we can decompose the infinitesimal deformation
vector field $\delta X^{\bar{\mu}}$ with respect to the spacetime
basis $\{ e_{a}, n^{I}\}$ as
\begin{equation}
\delta X^{\bar{\mu}}= \phi^{a} e{^{\bar{\mu}}}_{a}+
\phi{^{I}}n{^{\bar{\mu}}}_{I},
\end{equation}
therefore, we can find that the deformations of the intrinsic
geometry of the embedding are given by \cite{7,8,9,17}
\begin{equation}
\textbf{D}e_{a}= (K{_{ab}}^{I} \phi_{I})e^{b} + (\widetilde
\nabla_{a}\phi_{I}) n^{I} + (\nabla_{a} \phi^{b})e_{b} -
K{_{ab}}^{I} \phi^{b} n_{I},
\end{equation}
\begin{equation}
\textbf{D} \Gamma_{ab}= 2K{_{ab}}^{J} \phi_{J} + \nabla_{a}
\phi_{b}+\nabla_{b} \phi_{a},
\end{equation}
\begin{equation}
\textbf{D}\Gamma^{ab}= -2K^{abJ} \phi_{J} - \nabla^{a}
\phi^{b}-\nabla^{b} \phi^{a},
\end{equation}
\begin{equation}
\textbf{D}\sqrt{- \Gamma}= \sqrt{- \Gamma}[\nabla_{a}\phi^{a}+
K^{I} \phi_{I}],
\end{equation}
finally, the deformation of the extrinsic curvature is given by 
\begin{equation}
\textbf{D} K{_{ab}}^I= \widetilde \nabla_{a} \widetilde \nabla_{b} \phi^I + K{_{ac}} ^I K {^{c}} _{bJ} \phi^{J} + R((e_{a},n_{J}),e_{b},n^{I})\phi^{J}.
\end{equation}
where $R((e_{a},n_{J}),e_{b},n^{I})= R_{\bar{\alpha} \bar{\beta} \bar{\mu} \bar{ \nu}} e_{a}^{\bar{\beta}}  n_{J}^{\bar{\alpha}} e_{b}^{\bar{\mu}}  n^{I {\bar{\nu}}} $.  $R_{\bar{\alpha} \bar{\beta} \bar{\mu} \bar{ \nu}}$  is the riemann  tensor of the spacetime  with metric $g_{\bar{\mu}\bar{\nu}}$.  For this work these are all the  deformations that we will  need.\\
\noindent \textbf{ III. The Poincar\'e charges for chiral membranes by means of symplectic geometry   }\\[1ex]
In this section, we will find the Poincar\'e charges and the relevant Poisson brackets for the theory under study. In the literature \cite{12,13} we can find an equivalence between the chiral membrane dynamics and the DNG dynamics in an extended background spacetime plus a chirality condition,  we can see this equivalence  if  we considered   an action which is invariant under reparametrizations of the worldvolume  like DNG  with the induced metric  $\Gamma_{ab}$ 
\begin{equation}
S=-\mu_{0}\int d^{d+1} \xi \sqrt{-\Gamma},
\end{equation}
here $\Gamma$ is the determinant of the extended  induced metric $\Gamma_{ab}$,  $\mu_{0}$ is a constant  and the determinant  is given by $\Gamma= \gamma(1+ g_{44}\varpi)$,  $\gamma $ is the determinant of the induced metric $\gamma_{ab}$ without the field $\varphi$. In this manner, we can observe from action (16) that is one for superconducting strings involving  the Nielsen model whit $L(\varpi)= \sqrt{1+g_{44}}\varpi$ \cite{14}. Thus, with the embedding (1) we have unified the  strings and superconducting string theory.
In the next  lines,  we will find the covariant Poincar\'e charges  and we will establish the   covariant  quantization aspects  of the theory under study.  \\
Such as we commented in previous lines,  in recent works   using a covariant description of the canonical formalism for quantization  has been established the basis for the study of the symmetries and the quantization aspects of many systems such as DNG p-branes,  DNG p-branes  with thickness and  the topological strings governed for the Gauss-Bonnet  action \cite{7,8,9}. Such basis consists in constructing a covariant and gauge invariant symplectic structure on the corresponding quotient phase space $Z$   instead of choosing a special coordinate system on the phase space, with coordinates $p_i$ and $q^i$ as we usually find in the literature. \\
In view of    the action for our system is like  to DNG action,  for our purposes   we can use the results found in \cite{7, 8}  and the deformation formalism given in \cite{17}  to  construct our simplectic structure.\\
We follow the steps given in \cite{7, 8} first we will find the equations of motion,    which   determinate the covariant phase space defined as: {\it the space of solutions to the classical equations of motion} \cite{8,9, 15}. Using the action (17) and the deformations (11-15) we find 
\begin{equation}
\delta S=-\mu_{0}\int d^{d+1}\xi \sqrt{-\Gamma} \Gamma^{ab}K{_{ab}}^{I}\phi_{I} + \mu_0 \int d^{d+1} \nabla_a (\sqrt{-\Gamma} \phi^a) ,
\end{equation}
where we can identify the equations of motion
\begin{equation}
\Gamma^{ab} K_{ab}^{I}=0,
\end{equation}
and the divergence term $\Psi ^a=\sqrt{-\Gamma} \phi^a$ we can identified  as a simplectic potential,  because    the variation  of this potential will generate the simplectic structure for our theory \cite{7}.\\
On the other hand, if we take the variation of (18), we obtain the linearized equations, this is 
\begin{equation}
 \widetilde \Delta \phi^{I}+ K{_{ac}}^{I}K{^{ac}}_{J} \phi^{J}+
 R((e_{a},n_{J}),e_{a},n^{I})\phi^{J}=0,
\end{equation}
which has the same form  like  DNG for bosonic p-branes \cite{8, 9, 17}. \\
From the equation (19) and using the self-adjoint operators method we can find  a conserved current  and with this current we can construct our simplectic structure. However, following  \cite{7} we can construct the  same simplectic structure taking the variation of  the simpletic potential,  this is 
\begin{eqnarray}
\omega & = & \int \delta (-\mu_0 \sqrt{-\Gamma} \phi^a )d_{\Sigma_{a}}  =\int \delta(-\mu_0\sqrt{-\Gamma}e{^{a}}_{\bar{\mu}}\tau_{a})\wedge \delta
X^{\bar{\mu}}  d \Sigma \nonumber \\ &=&\int  \delta(- \mu_0\sqrt{-\Gamma}e{^{a}}_{{\mu}}\tau_{a}) \wedge \delta X^\mu d \Sigma + \int \delta(-\mu_0g_{44} \sqrt{-\Gamma} \varphi,_a \tau^a ) \wedge \delta \varphi d \Sigma ,
\end{eqnarray}
where we have used  the equations (1), (3) and (10), here $\delta$, is the deformation operator that acts as exterior derivative on the phase space \cite{6,7} and $\tau^a$ is a normalized ($\tau^a \tau_a =-1$) timelike vector field tangent to the world volume. It is remarkable to note  that $\omega$ given in (20) has the symmetries to be invariant under diffeomorphisms and world volume repametrizations, we can easy prove this in the same way that is presented in \cite{7}, also, $\omega$ turns out to be independent on the chose of $\Sigma$ {\it i.e.,} $\omega_\Sigma = \omega_{\Sigma'}$  where $\Sigma $ is a Cauchy d-surface. This property  will be important,  since it allows us to establish a connection  between functions and Hamiltonian vector fields on $Z$.\\
From (20) we can identify the canonical conjugate   coordinates, $p_{\mu} \equiv  -\mu_0 \sqrt{-\Gamma}e{^{a}}_{{\mu}}\tau_{a}$ is canonical conjugate to $q^{\mu} \equiv  X^\mu$ and  $ p' \equiv  -\mu_0g_{44} \sqrt{-\Gamma}\varphi,_a \tau^a$ to $q' \equiv \varphi$. In this manner, if $\varpi = 0$ we identify  the $p_\mu$ and  $X^\mu$ as the canonical variables for DNG bosonic membranes such as  is reported in \cite{8, 9}.\\
On the other hand, using equation (20)   we can group the canonical variables using only  one $\hat{P}^{\bar{\mu}}$  and $q^{\bar{\mu}}$,  this is 
\begin{equation}
\hat{P}_{\bar{\mu}}= -\mu_{0}\sqrt{-\Gamma} e{_{\bar{\mu}}}_{a} \tau^{a}, \quad
q^{\bar{\mu}}= X^{\bar{\mu}},
\end{equation}
and $\omega$ takes the form 
\begin{equation}
\omega= \int_{\Sigma} \delta \hat{P}_{\bar{\mu}} \wedge \delta X^{\bar{\mu}} d \Sigma.
\end{equation}
We can see  that the embedding  functions depend of local coordinates for the world-volume $(\xi^a)$  and we have that $\hat{P}_{\bar{\mu}} =\hat{P}^{\bar{\mu}} (\vec{\xi}, \tau)$,   $ X^{\bar{\mu}}=X^{\bar{\mu}} (\vec{\xi}, \tau)$, where we split the local coordinates for the world-volume in an arbitrary evolution parameter $\tau$ and coordinates $\vec{\xi}$ for $\Sigma$ at fixed values of  $\tau$. Thus, any function  $f$ on the phase space depends of $f=f(\hat{P}_{\bar{\mu}}, X^{\bar{\mu}})$.\\
Now, using basic ideas of symplectic geometry, we known that if  the symplectic structure $\omega$ is invariant under a group of transformations $G$, (in our case $\omega$ is invariant under  diffeormorphisms )  which corresponds to the gauge transformations of the theory \cite{7, 8}. Therefore, the Lie derivative along a vector $V$ tangent to a gauge orbit $G$ of $\omega$ vanishes, this is, 
\begin{equation}
\pounds_{V} \omega= V \rfloor \delta \omega + \delta (V \rfloor
\omega )= 0,
\end{equation}
where  $\rfloor$ denotes  the operation of contraction with $V$. Since $\omega$ is an exact and in particular closed two-form \cite{8,9} $\delta \omega=0$,   we have that  Eq. (23)  at least locally   takes the form  
\begin{equation}
V \rfloor \omega= -\delta H,
\end{equation}
where $H$ is a function on $Z$ which  we call the generator of the $G$ transformations. In this manner the relation (24) allows us to establish a connection between functions and Hamiltonian vector fields on $Z$.\\
On the other hand, if $h$ and $g$ are functions on the phase space, we can define using the simplectic structure $\omega$ a new function $[f,g]$, the Poisson bracket of $h$ and $g$, as 
\begin{equation}
[h,g]= V_{h} \rfloor g= - V_{g} \rfloor h,
\end{equation}
where $V_{h}$ and $V_{g}$ correspond to the Hamiltonian vector
fields generated by $h$, and $g$ respectively through Eq.
(23).\\
Now, we will calculate the fundamental Poisson brackets. For this, we take  the particular case of consider  the  flat spacetime ( $g_{\mu \nu} = \eta_{\mu \nu}$ see Eq. (2))  and using (24) we find 
\begin{eqnarray}
 \!\!\ & X^{\bar{\alpha}} & \!\!\ \quad \longrightarrow \quad V_{X^{\bar{\alpha}}}=
- \frac{\partial}{\partial p_{\bar{\alpha}}}, \nonumber \\
 \!\! & \hat{P_{\bar{\alpha}}}& \!\! \quad \longrightarrow \quad V_{P_{\bar{\alpha}}}=
\frac{\partial}{\partial X^{\bar{\alpha}}},
\end{eqnarray}
in this manner, using  (6) we have, 
\begin{equation}
[X^{\bar{\mu}}(\overrightarrow{\xi},
\tau),X^{\bar{\nu}}(\overrightarrow{\xi}',
\tau)]=[\hat{P}^{\bar{\mu}}(\overrightarrow{\xi},
\tau),\hat{P}^{\bar{\nu}}(\overrightarrow{\xi}', \tau)]=0,
\end{equation}
\begin{equation}
[X^{\bar{\mu}}(\overrightarrow{\xi},\tau),
\hat{P}_{\bar{\nu}}(\overrightarrow{\xi}', \tau) ]=
\delta{^{\bar{\mu}}}_{\bar{\nu}}\delta(\overrightarrow{\xi}-\overrightarrow{\xi}'),
\end{equation}
where $\delta{^{\mu}}_{\nu}$ is the Kronecker symbol and
$\delta(\overrightarrow{\xi}-\overrightarrow{\xi}') $  the Dirac
delta function. We can see from last equation that   in particular we have the following cases, for example;  $ [X^{\mu}(\overrightarrow{\xi},
\tau),X^{\nu}(\overrightarrow{\xi}',
\tau)]=[\hat{P}^{\mu}(\overrightarrow{\xi},
\tau),\hat{P}^{\nu}(\overrightarrow{\xi}', \tau)]=0, $ which correspond to  the Poisson brackets found in \cite{9} for [DNG] system, and  the cases; $[\varphi, \varphi]=[\varphi,_a \tau^a, \varphi,_b \tau^b]=0$, which corresponds to the Poisson brackets of the field  $\varphi$ living on the worldvolume.\\
On the other hand, if we choose  $V= \epsilon^{\bar{\alpha}}\frac{\partial}{\partial
X^{\bar{\alpha}}}$, where $\epsilon^ {\bar{\alpha}}$ is a constant in the Kaluza-Klein spacetime,  and  using (22),  (24)  we find  
\begin{equation}
V \rfloor \omega=- \delta
(-\epsilon_{\bar{\mu}}\tau_{a}(\mu_{0}\sqrt{- \Gamma}e^{a
\bar{\mu}})),
\end{equation}
where we can identify the linear momentum density 
\begin{equation}
P^{a \bar{\mu}}=-\mu_{0}\sqrt{- \Gamma}e^{a \bar{\mu}},
\end{equation}
using (6) we can prove that the lineal momentum for chiral membranes are covariantly conserved 
\begin{equation}
\nabla_{a} P^{a \bar{\mu}}= 0, 
\end{equation}
such as  is found in \cite{9, 18} for the [DNG] system. \\
We can express the total lineal momentum  $P^{\bar{\mu}}$ as 
\begin{equation}
P^{\bar{\mu}}= \int_\Sigma P^{a \bar{\mu}}d\Sigma_a= \int \hat{P}^{\bar{\mu}} d\Sigma,
\end{equation}
where $\hat{P}^{\bar{\mu}} $ is the canonical momentum. The total lineal momentum and the canonical momentum coincide.\\
Now, for a vector field given by $  V=\frac{a_{\bar{\alpha}\bar{\beta}}
}{2}[X^{\bar{\alpha}}\frac{\partial}{\partial
X^{\bar{\beta}}}-X^{\bar{\beta}} \frac{\partial}{\partial
X^{\bar{\alpha}}}]$, with $a_{\bar{\alpha}\bar{\beta}} = - a_{\bar{\beta} \bar{\alpha}} $,  the contraction $V \rfloor \omega $  gives
\begin{equation}
V \rfloor \omega=\delta (\frac{a_{\bar{\alpha}\bar{\beta}}}{2} P^{a
\bar{\mu}}X^{\bar{\nu}}- P^{a \bar{\alpha}}X^{\bar{\nu}}  ),
\end{equation}
thus, we can identify the angular momentum of the chiral membrane
\begin{equation}
M^{a \bar{\mu} \bar{\nu}}= \frac{1}{2}[P^{a
\bar{\mu}}X^{\bar{\nu}}- P^{a \bar{\alpha}}X^{\bar{\nu}}],
\end{equation}
using the gradients of the vector basis(Eq. (6)), we find that
\begin{equation}
\nabla_{a} M^{a \bar{\mu} \bar{\nu}}=0, 
\end{equation}
this is, the angular momentum is covariantly conserved too \cite{9, 18}.\\
We can define the total angular momentum $M ^{\bar{\alpha \bar{\beta}}}$  as 
\begin{equation}
M ^{\bar{\alpha} \bar{\beta}} = \int_{\Sigma}  M ^{ a\bar{\alpha} \bar{\beta}}  d\Sigma_a =\int_{\Sigma} (\hat{P}^{\bar{\beta} } X^{\bar{\alpha}} -  \hat{P}^{\bar{\alpha}} X^{\bar{\beta}} ) d\Sigma.
\end{equation}
In addition, we can use the equation (33) to find the Hamiltonian vector field associated to the angular momentum,  using (24) with $H^{\bar{\alpha}
\bar{\beta}}=(\hat{P}^{\bar{\beta}} X^{\bar{\alpha}}-
\hat{P}^{\bar{\alpha}} X^{\bar{\beta}}) $ we find
\begin{equation}
V^{\bar{\beta} \bar{\alpha}} = g^{ \bar{\lambda} \bar{\beta}}
\left( X^{\bar{\alpha} } \frac{\partial}{\partial
X^{\bar{\lambda}}} + \hat{P}^{\bar{\alpha}}
\frac{\partial}{\partial \hat{P}^{\bar{\lambda}}} \right)-
g^{\bar{\lambda} \bar{\alpha}} \left( X^{\bar{\beta}}
\frac{\partial}{\partial X^{\bar{\lambda}}}+ \hat{P}^{\bar{\beta}}
\frac{\partial}{\partial \hat{P}^{\bar{\lambda}}} \right),
\end{equation}
thus, using the definition of the Poisson's  brackets,  we can find $[M^{\bar{\mu}
\bar{\nu}}, M^{\bar{\alpha} \bar{\beta}}]$ and $[M^{\bar{\mu}
\bar{\nu}}, P^{\bar{\alpha}}]$, this is 
\begin{eqnarray}
\nonumber [M^{\bar{\mu} \bar{\nu}}, M^{\bar{\alpha} \bar{\beta}}]
\!\!\ & = & \!\!\ g^{\bar{\nu} \bar{\alpha}} M^{\bar{\mu}
\bar{\beta}} + g^{\bar{\mu} \bar{\alpha}} M^{\bar{\beta}
\bar{\nu}} + g^{\bar{\nu} \bar{\beta}}M^{\bar{\alpha}
\bar{\mu}} + g^{\bar{\mu} \bar{\beta}} M^{\bar{\nu} \bar{\alpha
}},
\end{eqnarray}
\begin{eqnarray}
 [M^{\bar{\mu} \bar{\nu}},P^{\bar{\alpha}} ] \!\!\ & = & \!\!\ g^{\bar{\mu} \bar{\alpha}} P^{\bar{\nu}}-
g^{\bar{\alpha} \bar{\nu}} P^{\bar{\mu}},
\end{eqnarray}
 We can see that the Poincar\'e charges,  $P$ and $M$,
indeed close correctly on the Poincar\'e algebra just  as the [DNG] theory \cite{9}, the difference with the [DNG] theory is that the Poison brackets given in  (38) contains the case for chiral membranes, and in particular  the results found in \cite{9}.\\

\noindent \textbf{ IV. Generalization of auxiliary variables for relativistic membranes }\\[1ex]
In this section, we will extend   the results given in \cite{16} for an arbitrary co-dimension and  we will find the stress tensor for the theory under study proving  that it coincide with the found in equation (30).\\
In  \cite{16}, has been considered a Hamiltonian which depends on the metric and extrinsic curvature induced on the surface. The metric and the curvature, along with the basis vector which connect them to the embedding functions defining the surface are introduced as auxiliary variables by adding appropriate constraints, all of them  quadratic. This elegant treatment, allow us study the response of the Hamiltonian to a deformations in each of the variables and the relationship between the multipliers implementing the constraints a the conserved stress tensor. \\
First, we will consider any reparametrization invariant functional of the variables $\Gamma_{ab}$ and $K{_{ab}}^I$ defined in the equations (3) and (7) respectively 
\begin{equation}
H[X]= \int \sqrt{-\Gamma} {\it H} d^{d+1} \xi.
\end{equation}
Such as in \cite{16},  we are interested in determining the response of $H$ to a deformation of the worldvolume $X \longrightarrow \delta X$, and will be to distribute the burden on $X$  among the basis normal $n^I$,  tangent $e^a$ and the extrinsic curvature $K{_{ab}}^I$ treating latter as independent auxiliary variables.\\
Now,  introducing  Lagrange multipliers function to implement the constraints we can  construct a new functional   $H_{c}[\Gamma_{ab},           
K_{ab}^{I},n^{I}, e_{a}, X^{\mu}, f^{a},
\Lambda{^{ab}}_{I}, \lambda^{ab},\lambda{^{a}}_{\perp I},
\lambda_{IJ}, \phi^a{_{IJ}} ] 
$    
\begin{eqnarray}
\nonumber H_{c}  \!\!\! & = & \!\!\ H[\Gamma_{ab}, K_{ab}^{I}]+
\int[f^{a} (e_{a}-
\partial_{a}X)+ \lambda{^{a}}_{\perp I} (e_{a}\cdot n^{I}) 
+ \lambda_{IJ}(n^{I}\cdot n^{I}
-\delta^{IJ})+ \Lambda{^{ab}}_{I} (K_{ab}^{I}- e_{a}\cdot\widetilde
\nabla_{b} n^{I})\nonumber \\
\!\!\ & + & \!\!\  \lambda^{ab}(g_{ab}-e_a\cdot e_{b})+
\phi{^{a}}_{IJ}(\omega{_{a}}^{IJ}- \nabla_{a}n^{J} \cdot n^{I}) ] \sqrt{-\Gamma} d^{d+1}\xi,
\end{eqnarray}
here, the original functional  is considered as a function of the independent variables $\Gamma_{ab}$  and $K{_{ab}}^I$, the auxiliary variables  $f^a$, $ \lambda{^{a}}_{\perp I} $, $ \lambda_{IJ}$, $ \Lambda{^{ab}}_{I} $, $\lambda^{ab}$, $\phi{^{a}}_{IJ}$ are Lagrange multipliers. In the expression (40) we  introduced  a new worldsheet covariant derivative $\widetilde
\nabla_{a}$ defined on fields transforming as tensors under frame
rotations as \cite{17}
\begin{equation}
\widetilde \nabla_{a} \phi{^{I}}_{J}= \nabla_{a} \phi{^{I}}_{J}-
\omega{_{a}}{^{I}}{_{k}}\phi{^{k}}_{J}-
\omega{_{a}}{_{J}}{^{k}}\phi{^{I}}_{k},
\end{equation}
the expression (41) and the Lagrange multiplier $\phi{^{a}}_{IJ}$,  are fundamentals expressions to extend   the results given in \cite{16} for  arbitrary co-dimensions.  \\
Using the equation (40) we can calculate the Euler-Lagrange equations  corresponding to $X$
\begin{equation}
\frac{\delta H_{c}}{\delta X}= \int \nabla_{a} (\sqrt{-\Gamma}f^{a})d^{D}\xi,
\end{equation}
where in the  equilibrium we have
\begin{equation}
\nabla_{a}(\sqrt{-\Gamma}f^{a})=0,
\end{equation}
this is,  $\sqrt{- \Gamma}f^{a}$ is covariantly conserved. The physical interpretation of $f^a$ is as a stress tensor  \cite{16}.\\
On the other hand, the Euler-Lagrange equations for $e_{a}$ are given by
\begin{equation}
\frac{\delta H_{c}}{\delta e_{a}}=\int[f^{a} +\lambda{^{a}}_{\perp
I}n^{I}- \Lambda{^{ab}}_{I} \nabla_{b}n^{I} +
\Lambda{^{ab}}_{I}\omega{_{b}}^{IJ}n_{J}- 2\lambda^{ab} e_{b}],
\end{equation}
in the equilibrium this implies 
\begin{equation}
f^{a} = -\lambda{^{a}}_{\perp I}n^{I}+ \Lambda{^{ab}}_{I}
\nabla_{b}n^{I} - \Lambda{^{ab}}_{I}\omega{_{b}}^{IJ}n_{J}+
2\lambda^{ab} e_{b},
\end{equation}
taking account   equation (41) we have  
\begin{equation}
f^{a}=-\lambda{^{a}}_{\perp I}n^{I} +\Lambda{^{ab}}_{I} \widetilde
\nabla_{b} n^{I}+ 2 \lambda^{ab}e_{b},
\end{equation}
using the Gauss-Weingarten equations  which describe completely the
extrinsic geometry of the worldsheet  Eq. (6), the last equation takes the form
\begin{equation}
f^{a}= (\Lambda{^{ab}}_{I} K{_{b}}^{cI}+2
\lambda^{ac})e_{c}-\lambda{^{a}}_{\perp I}n^{I},
\end{equation}
these expressions contains as particular case the found in \cite{16}.\\
In the same way, the Euler-Lagrange equations for  $n^I$ we have
\begin{eqnarray}
\nonumber \frac{\delta H_{c}}{\delta n^{I}} \!\!\ & = &\!\!\
\lambda{^{a}}_{\perp I} e_{a} +2 \lambda_{IJ} n^{J}+
\nabla_{a}(\Lambda{^{ab}}_{I}e_{b}) + \Lambda{^{ab}}_{I}
\omega{_{b}}^{IJ}e_{a}+ \nabla_{a}(\phi{^{a}}_{IJ}n^{J}) \nonumber
\\ \!\!\ & = & \!\! (\widetilde \nabla_{a} \Lambda{^{ab}}_{I}+ \lambda{^{b}}_{\perp
i}n^{I})e_{b} +(2 \lambda{^{J}}_{I}-\Lambda{^{ab}}_{I}K{_{ab}}^{J}
)n_{J}+ \nabla_{a}\phi{^{a}}_{IJ}n^{J} +
\phi{^{a}}_{IJ}\nabla_{a}n^{J},
\end{eqnarray}
and   the Euler-Lagrange equation for $\omega{_{a}}^{IJ}$ we find 
\begin{equation}
\frac{\delta H_{c}}{\delta \omega{_{a}}^{IJ}}=\Lambda{^{ab}}_{I}n_{J}\cdot e_{b}
+\phi{^{a}}_{IJ},
\end{equation}
from  the last equation we can see that in  equilibrium 
\begin{equation}
\phi{^{a}}_{IJ}=-\Lambda{^{ab}}_{I}n_{J}\cdot e_{b}.
\end{equation}
in this manner, enforcing the constraints (see Eqs. (5)) and using the expression (48) we have 
\begin{equation}
(\widetilde \nabla_{a} \Lambda{^{ab}}_{I}+ \lambda{^{b}}_{\perp
I})e_{b}+ (2\lambda{_{I}}^{J}
-\Lambda{^{ab}}_{I}K{_{ab}}^{J})n_{J}=0,
\end{equation}
this implies 
\begin{eqnarray}
\nonumber \widetilde \nabla_{a} \Lambda{^{ab}}_{I}+
\lambda{^{b}}_{\perp I}\!\!\ & = & \!\!\ 0 \nonumber \\
2\lambda{_{I}}^{J} -\Lambda{^{ab}}_{I}K{_{ab}}^{J}\!\!\ & = &
\!\!\ 0.
\end{eqnarray}
Finally, the Euler-Lagrange equations for $K{_{ab}}^{I}$ and $\Gamma_{ab}$ is given by 
\begin{eqnarray}
\nonumber \Lambda{^{ab}}_{I} \!\!\ & = & \!\!\ - \frac{\partial
H_{c}}{\partial K{_{ab}}^{I}} \nonumber \\
\lambda^{ab}\!\!\ & = & \!\!\ \frac{T^{ab}}{2},
\end{eqnarray}
where
\begin{equation}
T^{ab}= -2 (\sqrt{-\Gamma})^{-1} \frac{\partial \sqrt{- \Gamma
}H}{\partial \Gamma_{ab}}.
\end{equation}
Thus, to  study the case of chiral membranes we use  the expression (16) and    taking $H= -\mu_{0}$ in  the expression (54) we find
\begin{equation}
\lambda^{ab}=- \mu_{0} \Gamma^{ab}=2 T^{ab},
\end{equation}
using (55) in (47) we have
\begin{equation}
f^{a}=-\mu_{0} \Gamma^{ab} e_{b}= -\mu_{0}(\gamma^{ab}- g_{44}
\nabla^{a}\varphi \nabla^{b} \varphi)e_{b},
\end{equation}
where we can see that the stress tensor given in (56) coincides with the found in (30) using simplectic geometry.\\
Now, following the results given in \cite{16, 18}  we can decompose  the  stress $f^a$ in its normal and tangential parts, $f^{a}= F^{ab}e_{b}+
F{_{I}}^{a}n^{I}$, so the equations of motion are given by 
\begin{equation}
\nabla_{a}f^{a}= \widetilde \nabla_{a}f^{a}=0, 
\end{equation}
this is
\begin{eqnarray}
\nonumber \nabla_{a} F^{ab}+ K{^{bI}}_{a}F{_{I}}^{a} \!\!\ & = &
\!\!\ 0 \nonumber \\ \widetilde \nabla_{a}F^{aI}-
F^{ab}K{_{ab}}^{I} \!\!\ & = & \!\!\ 0.
\end{eqnarray}
For the   chiral membranes case we can see from  (56)  that $F^{aI}=0$, therefore,  the equations of motion (58)
takes the form
\begin{equation}
\Gamma^{ab}K_{ab}^{I}=0,
\end{equation}
this equations are the same found in Eq.(18). Thus, for I=i we have
\begin{equation}
(\gamma^{ab}- g_{44}\nabla^{a}\varphi \nabla^{b}\varphi)K{_{ab}}^{i}=0,
\end{equation}
and for I=4
\begin{equation}
\nabla_{a}\nabla^{a}\varphi=0,
\end{equation}
this  results coincide just  as is reported in \cite{13}.\\
\newline
\noindent \textbf{V. Conclusions and prospects}\\[1ex]
As we can see, in this paper   we have constructed a symplectic structure for chiral membranes  in an arbitrary co-dimension. With this geometric structure we could find the Poncar\'e charges, conservation laws , and construct the relevant Poisson brackets. In this manner, with these results we have the elements to study the quantization aspects in a covariant way. In particular, we can study the quantization aspects for  the case of chiral strings theory which is absent in the literature.\\
On the other hand, we extended  the results given in \cite{16} for an arbitrary co-dimension. Thus, with this extension we can reproduce important results found in the literature for example, conservation laws for several kind of extended objects that depends of the extrinsic geometry of the worldvolume  and   the stress tensor for objects of high-dimension  such as the systems   studied  in \cite{18, 19}. \\
\newline
\newline
\newline
\noindent \textbf{Acknowledgements}\\[1ex]
This work was supported by CONACYT under grant 44974-F.  The author wants to thanks R. Capovilla and Efrain Rojas for the support and the friendship that they  have offered me. \\
\newline
\newline
\newline
\newline


\begin{thebibliography}{}
\setlength{\itemsep}{-.50em}
\bibitem{1a} Elena Eizenberg and  Yuval Ne'eman, Membranes and Other Extendons (p-branes), Published 1995 World Scientific. 
\bibitem{1} E. Witten,  Nucl, Phys B 249 (1985) 537. 
\bibitem{2} S. C. Davis, A. C. Davis and M. Trodden, Phys. Lett. B 405 (1997) 257.
\bibitem{3} B. Carter and P. Peter, Phys, Lett. B 466 (1999) 41.
\bibitem{4} D. A. Steer, Phys. Rev. D 63 (2001) 083517: D. A. Steer, Strings Afther D term Inflation: Evolution and Properties of Chiral Cosmic Strings, astro-ph/0010295.
\bibitem{5} B. Carter, in formation and Interaction of Topological Defects (Nato ASI B349), Newton Institute, Cambridge, 1994, ed. R. Brandenberger and  A. C. Davis, pp 303 (Plenum, New York, 1995).
\bibitem{6} R. Cartas-Fuentevilla, Class. Quantum Grav., 19, 3571
(2002).
\bibitem{7} A. Escalante. Mod, Phys, Lett A, (2004), Vol 46, No 43, pp: 1902-1512.
\bibitem{8} A. Escalante, Int, J. Theor. Phys.  (2004), Vol 43, No.6, pp: 1491-1502.
\bibitem{9} A. Escalante, {\it Covariant Canonical formalism for Dirac-Nambu-Goto p-branes and
the Gauss-Bonnet topological term in string theory},  to be published in 
International Journal of Modern Physics  A, (2006).  
\bibitem{10} R. Cartas Fuentevilla and A. Escalante,{\it Topological terms and the global symplectic geometry of the
phase space in string theory}, Advances in Mathematical Research,
Nova Publishing (2004).
\bibitem{11} A. Polyakov, Nucl. Phys B 268 (1986), 406; H.
Kleinert, Phys. Lett B 174 (1986), 335.
\bibitem{12}  Ruben Cordero, Efrain Rojas. Int.J.Mod.Phys. A (2002) 17:73-88.
\bibitem{13} Ruben Cordero, Efrain Rojas. Rev. Mex.Fis. (2003) 49S1: 44-48, .
\bibitem{14a} J. J. Blanco-Pillado, Ken D. Olum, A. Vilenkin. Phys. Rev. D, (2001) 103513. 
\bibitem{14} N. K. Nielsen, Nucl, Phys. B 167 (1980) 249, 
\bibitem{15} C. Crncovi\'c and E. Witten, in {\it Three Hundred Years
of Gravitation}, edited by S. W. Hawking and W. Israel (Cambridge
 University Press. Cambridge, 1987).
\bibitem{16}  J. Guven,  J. Phys. A. (2004), 37:L313-L320.
\bibitem{17} R. Capovilla and J. Guven, Phys.\ Rev.\ D 51,
6736 (1995).
\bibitem{18} G. Arreaga, R. Capovilla, J. Guven, Ann. Phys. 279
(2000)126.
\bibitem{19}  R. Capovilla, J. Guven, E. Rojas. Class. Quant. Grav. 21:5563-5586, (2004).



\end{thebibliography}
\end{document}